\pdfoutput=1

\documentclass[11pt,twoside,a4paper,cmspaper,final,collab]{cms-tdr}
\def\svnVersion{174610}\def\cmsCernNo{2012-231}\def\cmsCernDate{\today}\def\cmsMessage{Submitted to Physical Review Letters}

\begin{document}\cmsNoteHeader{HIN-11-009}

\hyphenation{had-ron-i-za-tion}
\hyphenation{cal-or-i-me-ter}
\hyphenation{de-vices}

\RCS$Revision: 157952 $
\RCS$HeadURL: svn+ssh://svn.cern.ch/reps/tdr2/papers/HIN-11-009/trunk/HIN-11-009.tex $
\RCS$Id: HIN-11-009.tex 157952 2012-11-14 22:42:17Z alverson $
\providecommand{\HYDJET}{\textsc{hydjet}\xspace}
\cmsNoteHeader{HIN-11-009}
\title{Measurement of the azimuthal anisotropy of neutral pions in PbPb collisions at \texorpdfstring{$\sqrt{s_{NN}}$ = 2.76\TeV}{sqrt(s[NN])=2.76 TeV}}

\date{\today}

\abstract{
First measurements of the azimuthal anisotropy of neutral pions produced in PbPb collisions at a center-of-mass energy of $\sqrt{s_{NN}} = 2.76\TeV$ are presented. The amplitudes of the second Fourier component ($v_{2}$) of the $\Pgpz$ azimuthal distributions are extracted using an event-plane technique. The values of $v_{2}$ are studied as a function of the neutral pion transverse momentum (\pt) for different classes of collision centrality in the kinematic range $1.6 < \pt < 8.0\GeVc$, within the pseudorapidity interval $|\eta| < 0.8$. The CMS measurements of $v_{2}(\pt)$ are similar to previously reported \Pgpz\ azimuthal anisotropy results from $\sqrt{s_{NN}} = 200\GeV$ AuAu collisions at RHIC, despite a factor of  ${\sim}14$ increase in the center-of-mass energy. In the momentum range $2.5 < \pt < 5.0\GeVc$, the neutral pion anisotropies are found to be smaller than those observed by CMS for inclusive charged particles.
}

\hypersetup{%
pdfauthor={CMS Collaboration},%
pdftitle={Measurement of the azimuthal anisotropy of neutral pions in PbPb collisions at sqrt(s(NN)) = 2.76 TeV},%
pdfsubject={CMS},%
pdfkeywords={CMS, physics}}

\maketitle 

A central goal of relativistic heavy-ion experiments is to create a deconfined phase of nuclear matter, the quark gluon plasma (QGP), at extreme temperatures and energy densities, and to characterize its properties. Observations at the Relativistic Heavy Ion Collider (RHIC) suggest that an extremely dense partonic medium with near-perfect fluid properties is formed~\cite{PHENIX1,STAR,PHOBOS,BRAHMS}. These observations include the suppression of high-transverse-momentum (\pt) hadron production, referred to as ``jet-quenching"; strong azimuthal anisotropies in bulk particle production at low \pt; and baryon-meson differences in hadron suppression patterns and azimuthal anisotropies at intermediate \pt. Measurements of the azimuthal correlations of the produced particles play a key role in understanding the dominant physics processes in each of these transverse momentum ranges.

At low \pt (${<}2\GeVc$), the azimuthal anisotropy of the emitted particles is understood to be the result of a collective hydrodynamic expansion of the medium, converting any initial-state spatial anisotropy (eccentricity of the nuclear overlap region) into a final-state momentum anisotropy \cite{Tool1,Tool2}. The strength of the anisotropy is characterized by the values of the Fourier coefficients, $v_{n}$, of the expansion of the particle yields given by $\frac{\rd N}{\rd\phi_{R}} \propto 1 + \sum_n 2v_{n} \cos n(\phi - \psi_\mathrm{EP})$, where $\phi$ is the azimuthal angle of the outgoing particles and $\psi_\mathrm{EP}$ is the event plane angle reconstructed using the beam direction, $z$, and the azimuthal direction of the maximum transverse energy in each event. The second Fourier coefficient, $v_{2}$, is referred to as elliptic flow. At higher transverse momentum ($\pt \gtrsim 6\GeVc$), the azimuthal anisotropies have been attributed to the path-length dependence of energy loss in the medium due to the asymmetry in the reaction zone~\cite{gyulassy-2001-86,PhysRevC.80.054907,PHENIX,Bass:2008rv,CMSHighpT:2012}. In the intermediate \pt region, the RHIC data show an enhancement of baryon production~\cite{ppiPHENIX,Baryon1} and a larger $v_{2}$ of baryons as compared to mesons~\cite{BMratiosSTAR,PHENIX_scaling}. This behavior has been interpreted as a signature of quark recombination as the dominant production mechanism of moderate \pt hadrons, which implies the existence of quark degrees of freedom in the medium produced at RHIC~\cite{Muller200584}. Recent theoretical calculations also show that the RHIC measurements of baryon and meson $v_{2}$ at low \pt can be described by model calculations based on thermal partons only. However, contributions from shower partons  that are larger for mesons than for baryons must be included to explain the
data in the intermediate \pt range \cite{Hwa:2008um}.

The measurements of the elliptic anisotropy for inclusive charged particles, produced in PbPb collisions at a nucleon-nucleon center-of-mass energy of $\sqrt{s_{NN}}=2.76\TeV$ at the Large Hadron Collider (LHC), have been reported by ALICE \cite{ALICE}, ATLAS \cite{ATLAS}, and CMS \cite{CMS}. The measured $v_2$ coefficients exhibit similar strength and \pt dependence as those measured at RHIC in AuAu collisions \cite{ALICE,STARlowEnergy}. Differences in the elliptical flow of baryons and mesons can be indirectly tested via the comparison of the strength of the $v_2$ signals for $\Pgpz$ mesons and inclusive charged particles.

This Letter presents the first measurement of elliptic flow of $\Pgpz$ mesons as a function of \pt in PbPb collisions at a center-of-mass energy of $\sqrt{s_{NN}}= 2.76\TeV$. The data were recorded by the CMS experiment during the first LHC heavy-ion run in November 2010. The $\Pgpz$ meson elliptic flow is measured in the pseudorapidity range $|\eta|<$ 0.8, where $\eta$ is defined as $\eta  =  - \ln \left[ {\tan \left( {\theta /2} \right)} \right]$, and $\theta$ is the polar angle between the particle momentum and the anticlockwise beam direction. The measurement is performed over the full azimuthal coverage $0 < \phi < 2\pi$, and spans the range $1.6 < \pt < 8.0$\GeVc.

The detectors used for this analysis are the barrel Electromagnetic Calorimeter (ECAL) and the Hadron Forward (HF) calorimeter, which have an $\eta$ acceptance of $|\eta| < 1.4$ and $2.9 < |\eta| < 5.2$, respectively. Despite a wider pseudorapidity coverage of the barrel ECAL, these results are restricted to $|\eta| < 0.8$ in order to allow a direct comparison with the charged particle elliptic flow results \cite{CMS}. The barrel ECAL is located within a 3.8\unit{T} solenoidal magnetic field. The ECAL is made of lead-tungstate crystals that have a short radiation length (0.89\unit{cm}), and a small Moli\`{e}re radius (2.19\unit{cm}). A more detailed description of the CMS experiment can be found elsewhere \cite{CMSdetectors}.

The minimum-bias event sample is collected using coincidences between the trigger signals from each side of the interaction point using the Beam Scintillation Counters (BSC) ($3.23 < |\eta| < 4.65$) or the HF. Such a coincidence of minimum-bias trigger with bunches colliding in the interaction region suppresses any events due to noise, cosmic rays, out-of-time triggers, and beam backgrounds. The trigger accepts ($97\pm3$)\% of the total inelastic PbPb cross section. Collision centrality, defined as the fraction of total inelastic nucleus-nucleus cross section, is calculated using the sum of transverse energy ($\ET$) in towers from HF at both positive and negative $z$ positions \cite{CollCenCMS}. In this Letter, we present results based on centrality intervals of 10\% width, ranging from 20--30\% (more central) to 70--80\% (more peripheral). For the most central collisions (0--20\%), a small signal-to-background ratio limits the identification of $\Pgpz$ mesons.

The $\Pgpz$ mesons are measured by reconstructing their decay photons ($\Pgpz  \to \gamma \gamma$) in the barrel ECAL. Electromagnetic showers are found in the ECAL by forming clusters of contiguous crystals with a seed crystal having energy above a threshold of 200\MeV. Clusters are identified as photons on the basis of a shower shape requirement called the S4/S9 ratio. Photons are reconstructed using a $3 \times 3$ array of crystals, which contain on average 93\% of the photon energy. The quantity S4 is the total energy in a $2 \times 2$ array (a sub-matrix of the $3 \times 3$ array) containing the crystal with the highest energy deposited and S9 is the total energy in the $3 \times 3$ crystal matrix. There are four possible $2 \times 2$ matrix combinations, and S4 is defined as the most energetic of these four combinations. Clusters with S9 $>$ 400\MeV and S4/S9 $>$ 0.87 are selected as photon candidates for $\Pgpz$ meson reconstructed invariant mass, $m_{\gamma _i \gamma _j}$ calculations.
The invariant mass of a photon pair ($\gamma_{i}$,~$\gamma_{j}$) as measured in the ECAL is calculated from the energies and positions of the clusters, as given by $m_{\gamma _i \gamma _j} = \sqrt {2E_i E_j (1 - \cos \theta _{ij})}$,
where $\theta_{ij}$ is the opening angle between the two photons. Candidate pairs are formed from each photon cluster in an event in a particular \pt bin for the $\Pgpz$ meson invariant mass calculation. A \pt-dependent opening angle requirement and a cluster pair separation cut are also applied to the $\Pgpz$ meson invariant mass distribution. Pairs are selected with $\theta _{ij}~>~\frac{a}{{\pt }} + \frac{b}{{\pt^2 }}$, where {\it a} and {\it b} are opening angle cut parameters obtained from a detailed {\PYTHIA 6.422} simulation \cite{pythia}. The values of the parameters {\it a} and {\it b} are $0.17$\GeVc and $-0.11$ $(\!\GeVc)^{2}$, respectively. Further, a photon pair is rejected if the separation of the two photon clusters (distance is calculated based on the $\eta$ and $\phi$ coordinates of the clusters and using 1.29 m radius of the ECAL) is less than a threshold distance at a certain \pt. The threshold distance between photon-clusters decreases monotonically from 15\unit{cm} at \pt $\approx 1.6$\GeVc to 5.0\unit{cm} at \pt $\approx 8.0$\GeVc. At sufficiently high \pt, photons from a nearly symmetric decay ($E_{\gamma_{1}} \approx E_{\gamma_{2}}$, where $E_{\gamma_{i}}$ is the energy of a photon) can produce showers in the calorimeter that are reconstructed as a single cluster. In CMS, this effect becomes apparent at $\pt > 8.0\GeVc$. Consequently, results presented here are restricted to $\pt < 8.0\GeVc$.

The $\Pgpz$ meson yields are extracted statistically by subtracting the combinatorial background from the $\Pgpz$ candidate invariant mass distribution. The combinatorial background is estimated and subtracted using an event mixing technique, which forms pairs from photon candidates in different events. Each photon candidate is combined with all other photon candidates in three other events. The mixing of events is performed within intervals of centrality, {\it z}-vertex position and the event-plane angle \cite{CMS} orientation to replicate the background from uncorrelated pairs. All selections applied to the combinations of same-event pairs are also applied to mixed-event pairs. Event mixing is done in six $z$-vertex intervals of width $\Delta z =$ 5.0\unit{cm} in the range $|z| < 15\unit{cm}$. Similarly, the event-plane angle \cite{CMS} is also divided into six intervals in the range $0 < \psi_\mathrm{EP} < \pi$. The event-plane angle is determined from the HF, with flattening and resolution correction factors applied as in \cite{CMS}. The $\Pgpz$ reconstruction efficiencies as a function of $\pt$, centrality, and event-plane are studied by embedding simulated $\Pgpz$ mesons in real events. A total of 100\unit{k} such events are analyzed, where each event has ten $\Pgpz$ mesons embedded with a flat $\pt$ and $\phi$ distribution over a range of $0.2 < \pt < 10.0$\GeVc and $|\eta|<1.0$ to avoid any edge effects. The results for $\Pgpz$ meson elliptic flow are corrected for the dependence of the reconstruction efficiency on \pt for all centralities. In addition an in-plane versus out-of-plane dependence is observed for the $\Pgpz$ meson reconstruction efficiency in more central collisions. Corrections for this effect range from 16\% ($1.6 < \pt < 2.0$\GeVc) to 6\% ($2.5 < \pt < 3.0$\GeVc) for the 20--30\% centrality interval. For higher \pt intervals in this centrality class, such $\phi$-dependent efficiency corrections are not needed. Similarly for the 40--50\% and 50--60\% centrality intervals, these corrections range from 7\% ($1.6 < \pt < 2.0$\GeVc) to 4\% ($2.5 < \pt < 3.0$\GeVc), while no $\phi$-dependent efficiency corrections are needed for the more peripheral events.

\begin{figure}[!htb]
\begin{centering}
  \includegraphics[width=\linewidth]{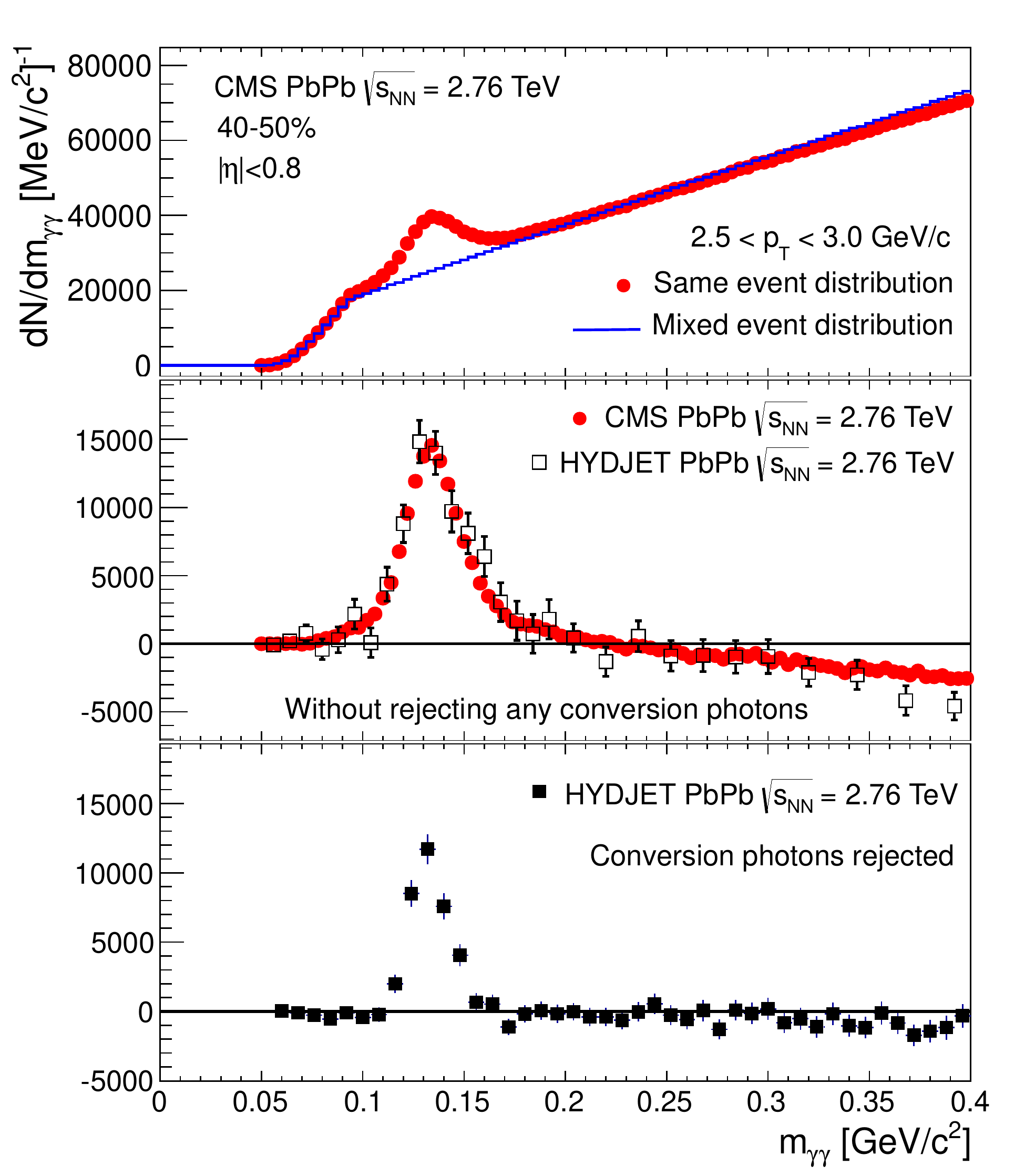}
  \caption{(Color online) Top panel: $\Pgpz$ meson invariant mass distribution with combinatorial background for $2.5<\pt<3.0\GeVc$ for 40--50\% centrality interval in PbPb collisions at $\sqrt{s_{NN}}=2.76\TeV$. The solid line represents the combinatorial background estimated from the event-mixing technique. Middle panel: Combinatorial background-subtracted $\Pgpz$ invariant mass distribution (solid circles) in the same $\pt$ bin and centrality interval. Open squares show HYDJET simulation results. Bottom panel: HYDJET simulation results obtained after rejecting converted photons. The HYDJET simulation results have been scaled up by a factor of 556 to match the data.}
  \label{fig:pi0InvMassDist}
  \end{centering}
\end{figure}

Figure~\ref{fig:pi0InvMassDist} (top panel) presents the $\Pgpz$ meson invariant mass distribution before background subtraction for $2.5 < \pt < 3.0\GeVc$ for the 40-50\% centrality interval. This panel shows the same-event distribution (solid circles) and the mixed-event normalized background (dashed line). For a given $\pt$ bin, the mixed-event background distribution is normalized to the same-event signal distribution in the range 200--250\MeVc. Different normalization regions such as 175--225\MeVc and 225--275\MeVc are also studied and no significant change in the resulting $\Pgpz$ meson $v_{2}$ is observed. The middle panel of Fig.~\ref{fig:pi0InvMassDist} shows the combinatorial-background-subtracted $\Pgpz$ meson invariant mass distribution (solid circles) in the same $\pt$ bin and centrality class. Over-subtraction is observed for higher mass regions, $m_{\gamma\gamma} >250\MeVc$. Investigations using \PYTHIA and \HYDJET (1.8) \cite{HYDJET} simulations show that this effect can be attributed to a correlated conversion background (converted photons) which has a different shape than a purely combinatorial background. By definition, the event-mixing technique cannot account for the effect of a correlated conversion background. Open symbols in the middle panel correspond to \HYDJET simulations, the result obtained without rejecting any converted photons. The background-subtracted mass spectrum predicted by simulations is seen to reproduce the data well. \HYDJET simulation results also show that the over-subtraction at high invariant mass is eliminated when the clusters from the converted photons are suppressed, as shown in the bottom panel of Fig. \ref{fig:pi0InvMassDist}. The event yield is calculated by integrating the data in a two standard deviations ($\sigma$, in units of mass) window around the mean ($\mu$) of the distribution. The $\sigma$ and $\mu$ are determined from a Gaussian fit to the combinatorial-background-subtracted $\Pgpz$ meson invariant mass distribution for every \pt and centrality interval. To avoid any model dependence, no corrections to the data are applied in order to account for these converted photons; instead asymmetric mass integration ranges of $\mu - 2\sigma < m_{\gamma\gamma} < \mu$, and $\mu - 3\sigma < m_{\gamma\gamma} < \mu$ are employed to understand the systematic effect of the conversion contribution to the mass peak in the higher mass regions. Studies showed that the maximum effect of the correlated background on the yield extraction in the mass integration range is less than 16\%.
To obtain the dependence of $\Pgpz$ meson production on azimuthal angle, the extracted yield is first measured in a given $\pt$ bin as a function of the azimuthal angle between the $\Pgpz$ meson trajectory and the event-plane orientation, $\psi_\mathrm{EP}$, found as described in Ref. \cite{CMS}. The measurement is performed in six equally spaced intervals of $\Delta \phi=\phi(\Pgpz) - \psi_\mathrm{EP}$ in the range $0 < \Delta \phi < \pi/2$. The $\Pgpz$ meson yields corrected for reconstruction efficiency are measured for each $\Delta \phi$ bin and the resulting angular distribution, $\rd N/\rd\Delta \phi$, is fitted with $N_0 \left( {1 + 2v_2 \cos 2\Delta \phi } \right)$ to determine the strength of the modulation in the yield.
We use an analytic linear $\chi^{2}$ fitting procedure that matches the integral of $N_0 \left( {1 + 2v_2 \cos 2\Delta \phi } \right)$ over each $\Delta \phi$ bin to the measured $\Pgpz$ meson yield within the corresponding bin \cite{BMratiosSTAR,PHENIX_scaling}.

Systematic uncertainties are assessed by varying the S4/S9 ratio and the mass integration ranges. A combination of the S4/S9 = 0.87 and $|m_{\gamma\gamma} - \mu| < 2.0\sigma$ mass integration range serves as a reference in this analysis. The $\Pgpz$ meson $v_{2}$ results are calculated for S4/S9 = 0.83 or 0.91 keeping the mass integration range at a reference value of $|m_{\gamma\gamma} - \mu| < 2.0\sigma$. In addition to asymmetric mass integration ranges, symmetric ranges such as $|m_{\gamma\gamma} - \mu| < 3.0\sigma$, and $|m_{\gamma\gamma} - \mu| < 1.5\sigma$ are used to determine the $\Pgpz$ meson $v_{2}$ results for all centralities keeping S4/S9 fixed at 0.87. The largest observed differences in the $v_{2}$ results based on different S4/S9 ratio cuts and $m_{\gamma\gamma} - \mu$ ranges are used to determine the systematic uncertainty. The systematic uncertainty determined from the precision of the $\phi$-efficiency curves obtained from the embedding procedure ranges from 18\% to 4\% from the lowest to the highest \pt intervals for 20--30\% centrality. For 70--80\% centrality, the systematic uncertainty varies from 7.2\% to 9\%.
The total systematic uncertainties obtained upon adding all the sources listed above in quadrature vary from 21\% ($1.6 < \pt < 2.0$\GeVc) to 31\% ($6.0 < \pt < 8.0$\GeVc) for the 20--30\% centrality interval. Similarly for 70--80\% these uncertainties change from 9.6\% ($1.6 < \pt < 2.0$\GeVc) to 33\% ($6.0 < \pt < 8.0$\GeVc). Systematic uncertainties arising from the trigger efficiency are found to be negligible.

\begin{figure}[!htb]
  \includegraphics[width=\linewidth]{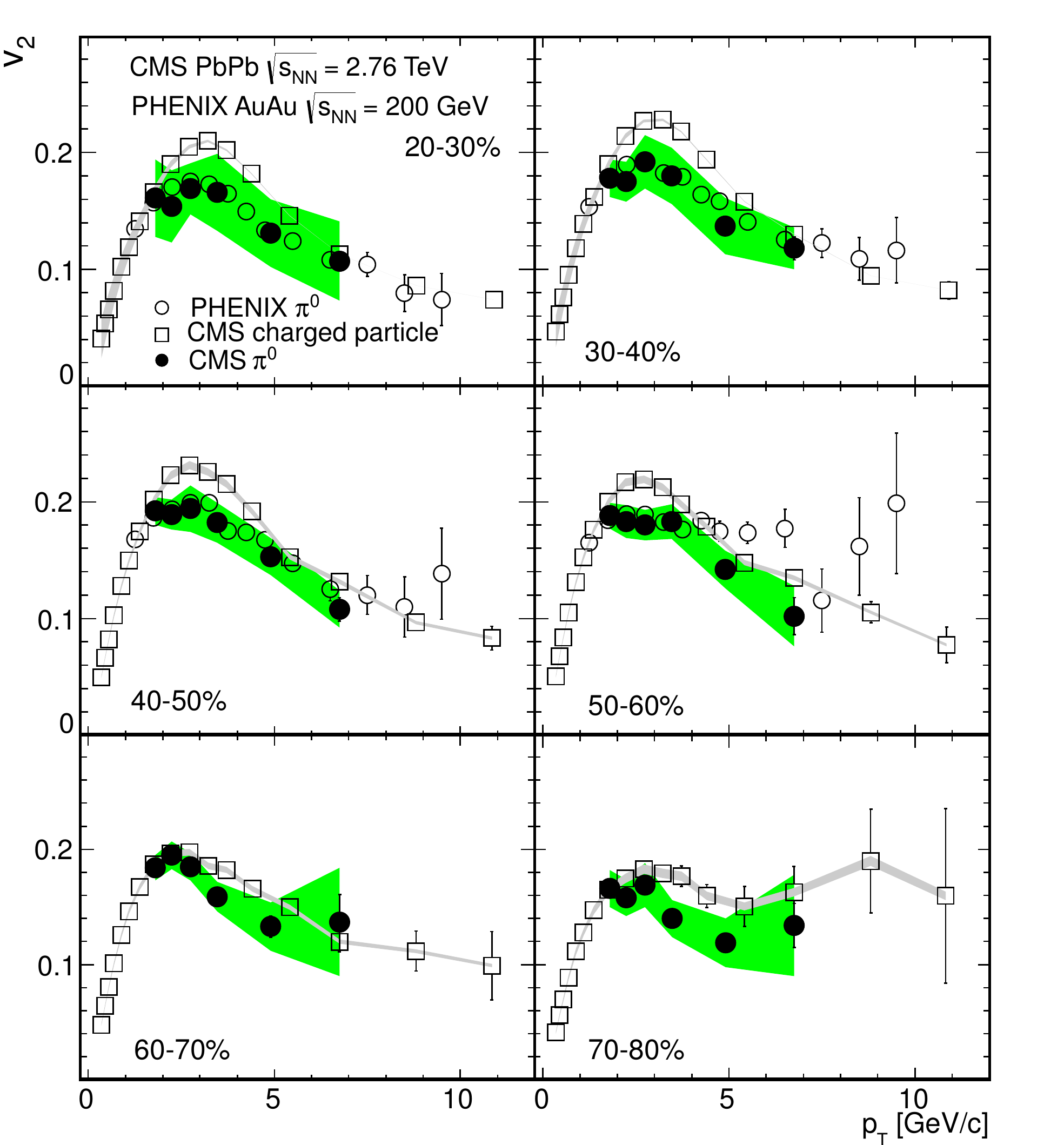}
  \caption{(Color online) CMS $\Pgpz$ meson $v_{2}$ (solid circles) compared to PHENIX $\Pgpz$ meson $v_{2}$ \cite{PHENIX} (open circles) for mid-rapidity ($|\eta| < 0.8$ and $|\eta| < 0.35$, respectively) and CMS charged particle $v_{2}$ (open squares, $|\eta| < 0.8$). Results are presented as a function of $\pt$ for six centrality intervals (20--30\% to 70--80\%). Green (grey) shaded bands represent systematic uncertainties associated with CMS $\Pgpz$ meson (charged particle) $v_{2}$ measurements. Only statistical uncertainties are shown for the PHENIX results. The systematic uncertainties for the 50--60\% centrality on the PHENIX data points are 10.4\%.}
  \label{fig:pi0v2comparison}
\end{figure}

The $\Pgpz$ meson $v_{2}(\pt)$ results are shown in Fig. \ref{fig:pi0v2comparison} for six centrality classes from 20--30\% to 70--80\%. CMS $\Pgpz$ meson $v_{2}$ results, shown as solid circles, are compared to PHENIX $\Pgpz$ $v_{2}$ results \cite{PHENIX}, for AuAu collisions at $\sqrt{s_{NN}}=200\GeV$, shown as open circles. Green (grey) shaded bands show the systematic uncertainties associated with the CMS $\Pgpz$ meson (charged particle) $v_{2}$ measurements. Our measurement shows qualitatively similar features as observed at RHIC energies despite an order of magnitude increase in the center-of-mass energy and the corresponding larger contribution from hard-scattered partons to meson production~\cite{Hwa:2011bw}. This observation is consistent with the previously reported similarity in the elliptic flow results for inclusive charged particles at RHIC and LHC \cite{ALICE,ATLAS}.

Figure \ref{fig:pi0v2comparison} also presents a comparison between CMS $\Pgpz$ meson $v_{2}$ results (solid circles), and CMS inclusive charged particle $v_{2}$ \cite{CMS} (open squares) as a function of $\pt$ using the event-plane method. The $\Pgpz$ meson $v_{2}$ is systematically lower than that for inclusive charged particles $v_{2}$ between $2.5 < \pt < 5.0$\GeVc for mid-central collisions (20--60\%). In more peripheral collisions (60-80\%), the differences tend to decrease for $\Pgpz$ mesons and inclusive charged particles, indicating a related origin of the elliptic anisotropy for all particle species.
For particles with intermediate \pt at RHIC, the $v_2$ values of baryons are observed to be higher than those for mesons~\cite{BMratiosSTAR,PHENIX_scaling}. The differences observed between the inclusive charged particle and $\Pgpz$ meson results may be due to the contribution from baryons which would increase the overall $v_{2}$ of the inclusive charged particles, compared to that for neutral pions, assuming a baryon-meson $v_{2}$ splitting comparable to that seen at RHIC. The baryon enhancement at RHIC has a
strong centrality dependence~\cite{ppiPHENIX,Baryon1,BMratiosSTAR}. Therefore, a detailed measurement of $v_{2}$ of identified particles as a function of centrality is important for understanding the production mechanism and the path-length dependence of parton energy loss in the medium.

In summary, the CMS detector has been used to perform the first measurements of the azimuthal anisotropy of neutral pions in PbPb collisions at $\sqrt{s_{NN}} = 2.76\TeV$. The measurements of $v_{2}$ were presented as a function of $\pt$ for six centralities,
from 20--30\% to 70--80\% for $1.6 < \pt < 8.0$\GeVc. Results were compared with PHENIX $\Pgpz$ meson \cite{PHENIX} and CMS inclusive charged particle measurements \cite{CMS}. It was found that the values of $v_{2}(\pt)$ for neutral pions measured at RHIC and
the LHC were of comparable magnitude. These data may shed light on the hadronization mechanism at RHIC and LHC energies, and contribute to the understanding of the parton-medium interactions. In the collision centrality interval 20--60\%, and in the momentum range $2.5<\pt<5.0\GeVc$, the magnitude of elliptic flow for neutral pions was found to be systematically lower than that for charged particles. This behavior is consistent with observations at lower collision energies, where this difference is found to be caused by the larger elliptic flow of baryons compared to mesons. The differences tend to decrease for more peripheral collisions (60--80\%) in the CMS data, where RHIC measurements are not available.

We congratulate our colleagues in the CERN accelerator departments for the excellent performance of the LHC machine. We thank the technical and administrative staff at CERN and other CMS institutes, and acknowledge support from: BMWF and FWF (Austria); FNRS and FWO (Belgium); CNPq, CAPES, FAPERJ, and FAPESP (Brazil); MES (Bulgaria); CERN; CAS, MoST, and NSFC (China); COLCIENCIAS (Colombia); MSES (Croatia); RPF (Cyprus); MoER, SF0690030s09 and ERDF (Estonia); Academy of Finland, MEC, and HIP (Finland); CEA and CNRS/IN2P3 (France); BMBF, DFG, and HGF (Germany); GSRT (Greece); OTKA and NKTH (Hungary); DAE and DST (India); IPM (Iran); SFI (Ireland); INFN (Italy); NRF and WCU (Korea); LAS (Lithuania); CINVESTAV, CONACYT, SEP, and UASLP-FAI (Mexico); MSI (New Zealand); PAEC (Pakistan); MSHE and NSC (Poland); FCT (Portugal); JINR (Armenia, Belarus, Georgia, Ukraine, Uzbekistan); MON, RosAtom, RAS and RFBR (Russia); MSTD (Serbia); SEIDI and CPAN (Spain); Swiss Funding Agencies (Switzerland); NSC (Taipei); TUBITAK and TAEK (Turkey); STFC (United Kingdom); DOE and NSF (USA).

\bibliography{auto_generated}   

\cleardoublepage \appendix\section{The CMS Collaboration \label{app:collab}}\begin{sloppypar}\hyphenpenalty=5000\widowpenalty=500\clubpenalty=5000\textbf{Yerevan Physics Institute,  Yerevan,  Armenia}\\*[0pt]
S.~Chatrchyan, V.~Khachatryan, A.M.~Sirunyan, A.~Tumasyan
\vskip\cmsinstskip
\textbf{Institut f\"{u}r Hochenergiephysik der OeAW,  Wien,  Austria}\\*[0pt]
W.~Adam, E.~Aguilo, T.~Bergauer, M.~Dragicevic, J.~Er\"{o}, C.~Fabjan\cmsAuthorMark{1}, M.~Friedl, R.~Fr\"{u}hwirth\cmsAuthorMark{1}, V.M.~Ghete, J.~Hammer, N.~H\"{o}rmann, J.~Hrubec, M.~Jeitler\cmsAuthorMark{1}, W.~Kiesenhofer, V.~Kn\"{u}nz, M.~Krammer\cmsAuthorMark{1}, I.~Kr\"{a}tschmer, D.~Liko, I.~Mikulec, M.~Pernicka$^{\textrm{\dag}}$, B.~Rahbaran, C.~Rohringer, H.~Rohringer, R.~Sch\"{o}fbeck, J.~Strauss, A.~Taurok, W.~Waltenberger, G.~Walzel, E.~Widl, C.-E.~Wulz\cmsAuthorMark{1}
% ...
\vskip\cmsinstskip
\dag:~Deceased\\
1:~~Also at Vienna University of Technology, Vienna, Austria\\
% ...

\end{sloppypar}
\end{document}